\documentclass[12pt]{article}
\usepackage[dvips]{graphicx}
\graphicspath{{figures/}}
\usepackage{subfigure}

\title{A Hadronization Model for Few-GeV Neutrino Interactions}
\author{T.~Yang \\  \it{Stanford University} \and C.~Andreopoulos \\ \it{Rutherford Appleton Laboratory}
\and H.~Gallagher, K.~Hofmann, P.~Kehayias \\ \it{Tufts University} }

\begin{document}
\maketitle
%
\abstract{
We present a detailed description of a new hadronic multiparticle production model for use in neutrino interaction simulations.
Its validity spans a wide invariant mass range starting from pion production threshold.
This model focuses on the low invariant mass region which is probed in few-GeV neutrino interactions and is of particular
importance to neutrino oscillation experiments using accelerator and atmospheric fluxes.
It exhibits reasonable agreement with a wide variety of experimental data.  We also describe measurements 
that can be made in upcoming experiments that can improve modeling in areas where uncertainties are currently large.  
%
\maketitle

\section{Introduction}
\label{intro}
In neutrino interaction simulations the hadronization model (or fragmentation model) determines the final 
state particles and 4-momenta given the nature of a neutrino-nucleon interaction (CC/NC, $\nu$/$\bar{\nu}$, target neutron/proton) 
and the event kinematics ($W^{2}$, $Q^{2}$, $x$, $y$).   
The modeling of neutrino-produced hadronic showers is important for a number of analyses 
in the current and coming generation of neutrino oscillation experiments:  

{\em Calorimetry:}  Neutrino oscillation experiments like MINOS which use calorimetry to reconstruct the shower energy, and hence the 
neutrino energy, are sensitive to the modelling of hadronic showers.  
These detectors are typically calibrated using single particle test beams, which introduces a model dependence in 
determining the conversion between detector activity and the energy of neutrino-produced hadronic systems  
\cite{Adamson:2007gu}.

{\em NC/CC Identification:}  Analyses which classify events as charged current (CC) or neutral current (NC) based on topological features such 
as track length in the few-GeV region rely on accurate simulation of hadronic particle distributions to determine
NC contamination in CC samples.

{\em Topological Classification:}  
Analyses which rely on topological classifications, for instance selecting quasi-elastic-like events
based on track or ring counting depend on the simulation of hadronic systems to determine
feeddown of multi-particle states into selected samples.  Because of the wide-band nature of most current neutrino beams, 
this feeddown is non-neglible even for experiments operating in beams with mean energy as low as 1 GeV 
\cite{miniBoone:2007ru,Hiraide:2008yu}.    

{\em $\nu_e$ Appearance Backgrounds:}  
A new generation of $\nu_{\mu}\rightarrow\nu_{e}$ appearance experiments are being developed around the world, which 
hope to measure $\theta_{13}$, resolve the neutrino mass hierarchy, and find evidence of CP violation in the lepton sector \cite{Hayato:2005hq,Ayres:2004js}.
In these experiments background is dominated by neutral pions generated in NC interaction. 
The evaluation of NC backgrounds in these analysis can be 
quite sensitive to the details of the NC shower simulation and specifically the $\pi^{0}$ shower content and transverse
momentum distributions of hadrons \cite{Sanchez:2008zza}.  

\begin{figure*}
\centering
  \includegraphics[width=\textwidth]{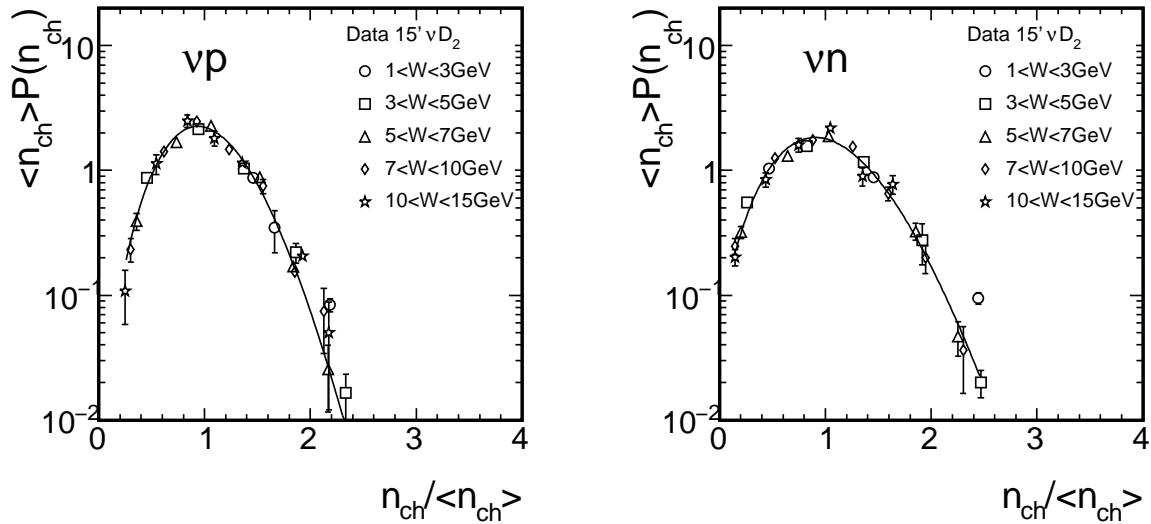}%
  \caption{KNO scaling distributions for $\nu p$ (left) and $\nu n$ interactions. The curve represents a fit to the Levy function. Data points are taken from \cite{Zieminska:1983bs}.}\label{fig:kno_levy}
\end{figure*}


In order to improve Monte Carlo simulations for the MINOS experiment, a new hadronization model, referred to here as the `AGKY model', was developed.
We use the PYTHIA/JETSET \cite{Sjostrand:2006za} model to simulate the hadronic showers at high hadronic invariant masses. We also developed a phenomenological description of the low invariant mass hadronization since the applicability of the \\
PYTHIA/JETSET model, for neutrino-induced showers, is known to deteriorate as one approaches the pion production threshold.   We present here a description of the AGKY hadronization model and the tuning and validation of this model using bubble chamber experimental data.

\section{The AGKY Model}
\label{sec:1}

\subsection{Overview}
\label{sec:2}

The AGKY model, which is now the default hadronization model in the neutrino Monte Carlo generators NEUGEN \cite{Gallagher:2002sf} and GENIE-2.0.0 \cite{Andreopoulos:2006cz}, 
includes a phenomenological description of the low invariant mass region based on Koba-Nielsen-Olesen (KNO) scaling \cite{Koba:1972ng},  
while at higher masses it gradually switches over to the PYTHIA/JETSET model. 
The transition from the KNO-based model to the  \\
PYTHIA/JETSET model takes place gradually, at an intermediate invariant mass region, ensuring the continuity of all simulated observables as a function of the invariant mass. This is accomplished by using a transition window [$W_{min}^{tr},W_{max}^{tr}$] over which we linearly increase the fraction of neutrino events for which the hadronization is performed by the PYTHIA/JETSET model from 0\% at $W_{min}^{tr}$ to 100\% at $W_{max}^{tr}$. The default values used in the AGKY model are:
\begin{equation}
W_{min}^{tr} = 2.3\mbox{ GeV}/\mbox{c}^{2}, W_{max}^{tr} = 3.0\mbox{ GeV}/\mbox{c}^{2}.  
\end{equation}


The kinematic region probed by any particular experiment depends on the neutrino flux, and for the 1-10 GeV range of importance to 
oscillation experiments, the KNO-based phenomenological description plays a particularly crucial role.  The higher invariant 
mass region where PYTHIA/JETSET is used is not accessed until a neutrino energy of approximately 3 GeV is reached, at which point 
44.6\% of charged current interactions are non-resonant inelastic and are hadronized using the KNO-based part of the model.  
For 1 GeV neutrinos this component is 8.3\%, indicating that this model plays a significant role even at relatively low 
neutrino energies.   At 9 GeV, the contributions from the KNO-based and PYTHIA/JETSET components of the model are approximately 
equal, with each handling around 40\% of generated CC interactions.  
The main thrust of this work was to improve the 
modeling of hadronic showers in this low invariant mass / energy regime which is 
of importance to oscillation experiments. 


The description of AGKY's KNO model, used at low invariant masses, can be split into two independent parts:
\begin{itemize}
\item Generation of the hadron shower particle content
\item Generation of hadron 4-momenta
\end{itemize}
These two will be described in detail in the following sections.

The neutrino interactions are often described by the following kinematic variables:
\begin{eqnarray}
Q^{2} &=& 2E_{\nu}(E_{\mu}-p_{\mu}^{L})-m^{2}\nonumber\\
\nu &=& E_{\nu} - E_{\mu}\nonumber\\
W^{2} &=& M^{2} + 2M\nu -Q^{2}\nonumber\\
x &=& Q^{2}/2M\nu\nonumber\\
y &=& \nu/E_{\nu}
\end{eqnarray}
where $Q^{2}$ is the invariant 4-momentum transfer squared, $\nu$ is the neutrino energy transfer, $W$ is the effective mass of all secondary hadrons (invariant hadronic mass), $x$ is the Bjorken scaling variable, $y$ is the relative energy transfer, $E_{\nu}$ is the incident neutrino energy, $E_{\mu}$ and $p_{\mu}^{L}$ are the energy and longitudinal momentum of the muon, $M$ is the nucleon mass and $m$ is the muon mass. 

For each hadron in the hadronic system, we define the variables $z=E_{h}/\nu$, $x_{F}=2p_{L}^{*}/W$ and $p_{T}$ where $E_{h}$ is the energy in the laboratory frame, $p_{L}^{*}$ is the longitudinal momentum in the hadronic c.m.s., and $p_{T}$ is the transverse momentum. 

\subsection{Low-$W$ model: Particle content}
\label{sec:3}
At low invariant masses the AGKY model generates hadronic systems that typically consist of exactly one baryon ($p$ or $n$) and any 
number of 
$\pi$ and K mesons that are kinematically possible and consistent with charge conservation.

For a fixed hadronic invariant mass and initial state (neutrino and struck nucleon), the method for generating the hadron shower particles generally proceeds in four steps:

{\em Determine $\langle n_{ch} \rangle$:}  Compute the average charged hadron multiplicity using the empirical expression:
\begin{equation} 
\langle n_{ch}\rangle = a_{ch} + b_{ch}\  \ln W^{2}
\end{equation}
The coefficients $a_{ch}$, $b_{ch}$, which depend on the initial state, have been determined by bubble chamber experiments.

{\em Determine $\langle n \rangle$:}  Compute the average hadron multiplicity as $\langle n_{tot}\rangle=1.5\langle n_{ch}\rangle$ \cite{Wittek:1988ke}.

{\em Deterimine n:} Generate the actual hadron multiplicity taking into account that the multiplicity dispersion is described by the KNO scaling law \cite{Koba:1972ng}:
\begin{equation}
\langle n \rangle \times P(n) = f(n/\langle n \rangle)
\end{equation}
where $P(n)$ is the probability of generating $n$ hadrons and $f$ is the universal scaling function which can be parametrized by the Levy function  
\footnote{The Levy function: $Levy(z;c) = 2 e^{-c}c^{c z + 1} /\Gamma(cz+1)$} 
($z=n/\langle n \rangle$) with an input parameter $c$ that depends on the initial state. Fig.\ref{fig:kno_levy} shows the KNO scaling distributions for $\nu p$ (left) and $\nu n$ (right) CC
interactions. 
We fit the data points to the Levy function and the best fit parameters are $c_{ch} = 7.93 \pm 0.34$ for the $\nu p$ interactions 
and $c_{ch} = 5.22 \pm 0.15$ for the $\nu n$ interactions.

{\em Select particle types:} Select hadrons up to the generated hadron multiplicity taking into account charge conservation and kinematic constraints. 
The hadronic system contains any number of mesons and exactly one baryon which is generated based on simple quark model arguments. 
Protons and neutrons are produced in the ratio 
2:1 for $\nu p$ interactions, 1:1 for $\nu n$ and $\bar{\nu}p$, and 1:2 for $\bar{\nu}n$ interactions.   
Charged mesons are then created in order to balance charge, and the remaining  mesons are generated in neutral pairs. The probablilities for each 
are  
31.33\% ($\pi^{0},\pi^{0}$), 62.66\% ($\pi^{+},\pi^{-}$), 1.5\% ($K^{0},K^{-}$), 
1.5\% ($K^{+},K^{-}$), 1.5\% ($\bar{K^{0}},K^{+}$) and 1.5\% ($K^{0},\bar{K^{0}}$).  
The probability of producing a strange baryon via associated production is determined from a fit to $\Lambda$ production data:
\begin{equation} 
P_{hyperon}= a_{hyperon}+ b_{hyperon}\  \ln W^{2}
\end{equation}

TABLE \ref{tab:par} shows the default average hadron multiplicity and dispersion parameters used in the AGKY model.


\begin{table}
\center
\caption{Default AGKY average hadron multiplicity and dispersion parameters (see text for details).}
\label{tab:par}
\begin{tabular}{|l|r|r|r|r|}
\hline

    &$\nu p$&$\nu n$&$\bar{\nu}p$&$\bar{\nu}n$\\
\hline
$a_{ch}$ & 0.40 \cite{Zieminska:1983bs} & -0.20 \cite{Zieminska:1983bs}& 0.02 \cite{Barlag:1981wu} & 0.80 \cite{Barlag:1981wu}\\
$b_{ch}$ & 1.42 \cite{Zieminska:1983bs} & 1.42 \cite{Zieminska:1983bs}& 1.28 \cite{Barlag:1981wu} & 0.95 \cite{Barlag:1981wu}\\
$c_{ch}$ & 7.93 \cite{Zieminska:1983bs} & 5.22 \cite{Zieminska:1983bs}& 5.22  & 7.93\\
$a_{hyperon}$ & 0.022 & 0.022 & 0.022  & 0.022\\
$b_{hyperon}$ & 0.042 & 0.042 & 0.042  & 0.042\\
\hline
\end{tabular}
\end{table}

\subsection{Low-$W$ model: Hadron system decay}
\label{sec:4}

Once an acceptable particle content has been generated, the available invariant mass needs to be partitioned amongst the generated hadrons. 
The most pronounced kinematic features in the low-$W$ region result from the fact that the produced baryon is much heavier than the mesons and 
exhibits a strong directional anticorrelation with the current direction.

Our strategy is to first attempt to reproduce the experimentally measured final state nucleon momentum distributions. 
We then perform a phase space decay on the remnant system employing, in addition, a $p_T$-based rejection scheme designed to reproduce the expected meson transverse momentum distribution.  The hadronization model performs its calculation in the hadronic c.m.s., where the z-axis is in the direction of the momentum transfer.  
Once the hadronization is completed, the hadronic system will be boosted and rotated to the LAB frame. The boost and rotation maintains the $p_{T}$ generated in the hadronic c.m.s. 

In more detail, the algorithm for decaying a system of $N$ hadrons is the following:

{\em Generate baryon:}  Generate the baryon 4-momentum $P_{N}^{*} = (E_{N}^{*},{\bf p}_{N}^{*})$ using the nucleon $p_{T}^{2}$ and $x_{F}$ PDFs which are parametrized 
based on experimental data \cite{Derrick:1977zi,CooperSarkar:1982ay}.
The $x_F$ distribution used is shown in Fig.\ref{fig:xf}. 
We do not take into account the correlation between $p_{T}$ and $x_{F}$ in our selection. 


\begin{figure}
\centering
    \includegraphics[width=0.5\textwidth]{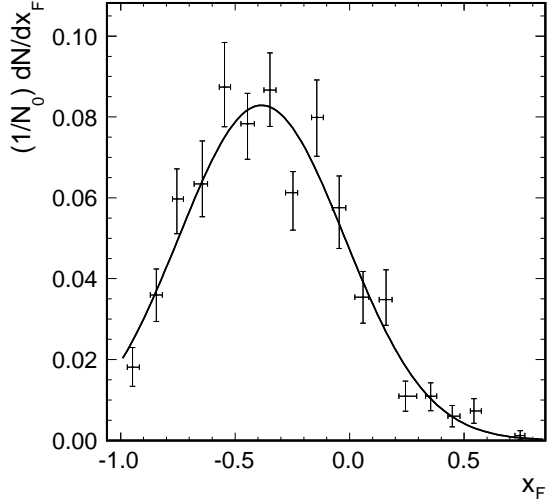}
\caption{\label{fig:xf}  Nucleon  $x_{F}$ distribution data from Cooper $et\ al.$ \cite{CooperSarkar:1982ay} and the AGKY parametrization (solid line).   
}
\end{figure}

{\em Remnant System:}  Once an accepted $P^{*}_{N}$ has been generated, calculate the 4-momentum of the remaining N-1 hadrons, (the ``remnant'' hadronic system) as $P_{R}^{*} = P_{X}^{*} - P_{N}^{*}$ where $P_{X}^{*} = (W,0)$ is the initial hadron shower 4-momentum in the hadronic c.m.s.

{\em Decay Remnant System:}  Generate an unweighted phase space decay of the remnant hadronic system \cite{James:1968}. The decay takes place at the remnant system c.m.s. 
after which the particles are boosted back to the hadronic c.m.s. 
The phase space decay employs a rejection method suggested in \cite{Clegg:1981js}, 
with a rejection factor $e^{-A*p_{T}}$ for each meson. This causes the transverse momentum distribution of the generated mesons to fall exponentially with increasing $p^{2}_{T}$.
Here $p_{T}$ is the momentum component perpendicular to the current direction. 

Two-body hadronic systems are treated as a special case.  Their decay 
is performed isotropically in the hadronic c.m.s. and no $p_{T}$-based suppression factor is applied.

\subsection{High-$W$ model: PYTHIA/JETSET}
\label{sec:5}
The high invariant mass hadronization is performed by the PYTHIA/JETSET model \cite{Sjostrand:2006za}. The PYTHIA program is a standard tool for the generation of high-energy collisions, comprising a coherent set of physics models for the evolution from a few-body hard process to a complex multihadronic final state. It contains a library of hard processes and models for initial- and final-state parton showers, multiple parton-parton interactions, beam remnants, string fragmentation and particle decays. The hadronization model in PYTHIA is based 
on the Lund string fragmentation framework \cite{Andersson:1983ia}. In the AGKY model, all but four of the PYTHIA configuration parameters are set to be the default values. Those four parameters take the non-default values tuned by NUX \cite{Rubbia:2001}, a high energy neutrino MC generator used by the NOMAD experiment:
\begin{itemize}
\item $P_{s\bar{s}}$ controlling the ${s\bar{s}}$ production suppression: \\ 
      (PARJ(2))=0.21.
\item $P_{\langle p_{T}^{2} \rangle}$ determining the average hadron $\langle p_{T}^{2} \rangle$: \\
      (PARJ(21))=0.44.
\item $P_{ngt}$ parameterizing the non-gaussian $p_{T}$ tails: \\
      (PARJ(23))=0.01. 
\item $P_{Ec}$ an energy cutoff for the fragmentation process: \\ 
      (PARJ(33))=0.20.
\end{itemize}

\section{Data/MC Comparisons}
\label{sec:6}

The characteristics of neutrino-produced hadronic systems have been extensively studied by several bubble chamber experiments. 
The bubble chamber technique is well suited for studying details of charged hadron production in neutrino interactions since the detector 
can provide precise information for each track.
However, the bubble chamber has disadvantages for measurements of hadronic system characteristics as well.    
The detection of neutral particles, in particular of photons from $\pi^{0}$ decay, 
was difficult for the low density hydrogen and deuterium experiments.    Experiments that measured neutral pions typically used 
heavily liquids such as neon-hydrogen mixtures and Freon.  While these exposures had the advantage of higher statistics and 
improved neutral particle identification, they had the disadvantage of introducing intranuclear rescattering which complicates
the extraction of information related to the hadronization process itself.  

We tried to distill the vast literature and focus on the following aspects of $\nu$/$\bar{\nu}$ measurements made in three bubble chambers - the Big European Bubble Chamber (BEBC) at CERN, the 15-foot bubble chamber at Fermilab, and the SKAT bubble chamber in Russia. 
Measurements from the experiments of particular interest for tuning purposes can be broadly categorized as multiplicity measurements and 
hadronic system measurements.  Multiplicity measurements include
averaged charged and neutral particle ($\pi^{0}$) multiplicities, 
forward and backward hemisphere average multiplicities and correlations,
topological cross sections of charged particles, and neutral - charged pion multiplicity correlations.
Hadronic system measurements include 
	fragmentation functions ($z$ distributions),
        $x_{F}$ distributions,
        $p^{2}_{T}$ (transverse momentum squared) distributions, and 
        $x_{F} - \langle p_{T}^{2} \rangle$ correlations (``seagull'' plots).

\begin{figure*}
\centering
  \includegraphics[width=\textwidth]{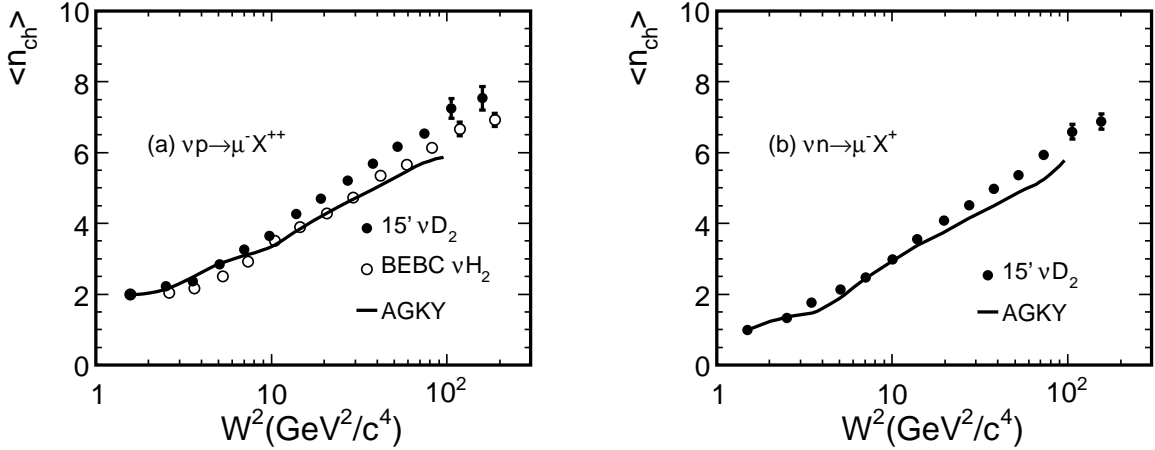}
  \caption{Average charged-hadron multiplicity $\langle n_{ch} \rangle$ as a function of $W^{2}$. (a) $\nu p$ events. (b) $\nu n$ events. Data points are taken from \cite{Zieminska:1983bs,Allen:1981vh}.}\label{fig:cMulCh}
\end{figure*}

The systematic errors in many of these measurements are substantial and various corrections had to be made to correct for muon selection efficiency, neutrino energy smearing, {\it etc}. The direction of the incident $\nu$/$\bar{\nu}$ is well known from the geometry of the beam and the position of the interaction point. Its energy is unknown and is usually estimated using a method based on transverse momentum imbalance. The muon is usually identified through the kinematic information or by using an external muon identifier (EMI). 
The resolution in neutrino energy is typically 10\% in the bubble chamber experiments and the invariant hadronic mass $W$ is less well determined.

The differential cross section for semi-inclusive pion production in neutrino interactions 
\begin{equation}
\nu + N \rightarrow \mu^{-}+\pi+X
\label{nuN}
\end{equation}
may in general be written as:
\begin{equation}
\frac{d\sigma(x,Q^{2},z)}{dxdQ^{2}dz} = \frac{d\sigma(x,Q^{2})}{dxdQ^{2}}D^{\pi}(x,Q^{2},z),
\end{equation}
where $D^{\pi}(x,Q^{2},z)$ is the pion fragmentation function. Experimentally $D^{\pi}$ is determined as:
\begin{equation}
D^{\pi}(x,Q^{2},z) = [N_{ev}(x,Q^{2})]^{-1}dN/dz.
\end{equation}

In the framework of the Quark Parton Model (QPM) the dominant mechanism for reactions (\ref{nuN}) is the interaction of the exchanged $W$ boson with a d-quark to give a u-quark which fragments into hadrons in neutrino interactions, leaving a di-quark spectator system which produces target fragments. In this picture the fragmentation function is independent of $x$ and the scaling hypothesis excludes a $Q^{2}$ dependence; therefore the fragmentation function should depend only on $z$. There is no reliable way to separate the current fragmentation region from the target fragmentation region if the effective mass of the hadronic system ($W$) is not sufficiently high. Most experiments required $W>W_{0}$ where $W_{0}$ is between 3 GeV/$c^{2}$ and 4 GeV/$c^{2}$ when studying the fragmentation characteristics. The caused difficulties in the tuning of our model because we are mostly interested in the interactions at low hadronic invariant masses. 

\begin{figure*}
\centering
  \includegraphics[width=\textwidth]{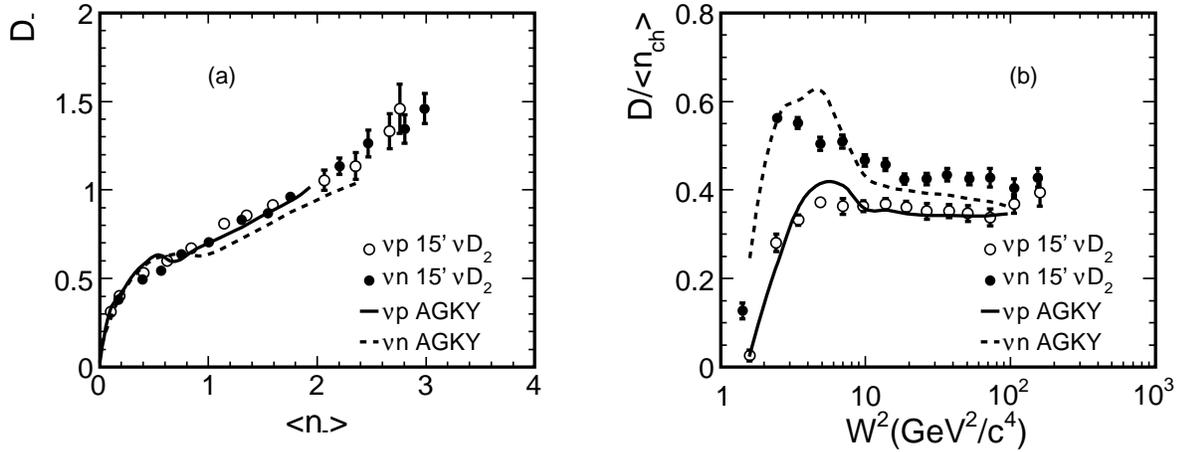}
  \caption{(a) The dispersion $D_{-} = (\langle n_{-}^{2} \rangle - \langle n_{-} \rangle ^{2})^{1/2}$ as a function of $\langle n_{-} \rangle$. (b) $D/\langle n_{ch} \rangle$ as a function of $W^{2}$. Data points are taken from \cite{Zieminska:1983bs}.}\label{fig:cDispCh}
\end{figure*}

We determined the parameters in our model by fitting experimental data with simulated CC neutrino free nucleon interactions uniformly distributed in the energy range from 1 to 61 GeV. 
The events were analyzed to determine the hadronic system characteristics and compared with published experimental data from the  
BEBC, Fermilab 15-foot, and SKAT bubble chamber experiments. 
We reweight our MC to the energy spectrum measured by the experiment if that information is available.  
This step is not strictly necessary for the following two reasons: many observables (mean multiplicity, dispersion, {\it etc.}) 
are measured as a function of the hadronic invariant mass $W$, in which case the energy dependency is removed; secondly the scaling variables ($x_{F}$, $z$, {\it etc.}) 
are rather independent of energy according to the scaling hypothesis. 

Some experiments required $Q^{2}>1\mbox{GeV}^{2}$ to reduce the quasi-elastic contribution, $y<0.9$ to reduce the neutral currents, and $x>0.1$ to reduce the sea-quark contribution. They often applied a cut on the muon momentum to select clean CC events. 
We apply the same kinematic cuts as explicitly stated in the papers to our simulated events.

\begin{figure*}
\centering
  \includegraphics[width=\textwidth]{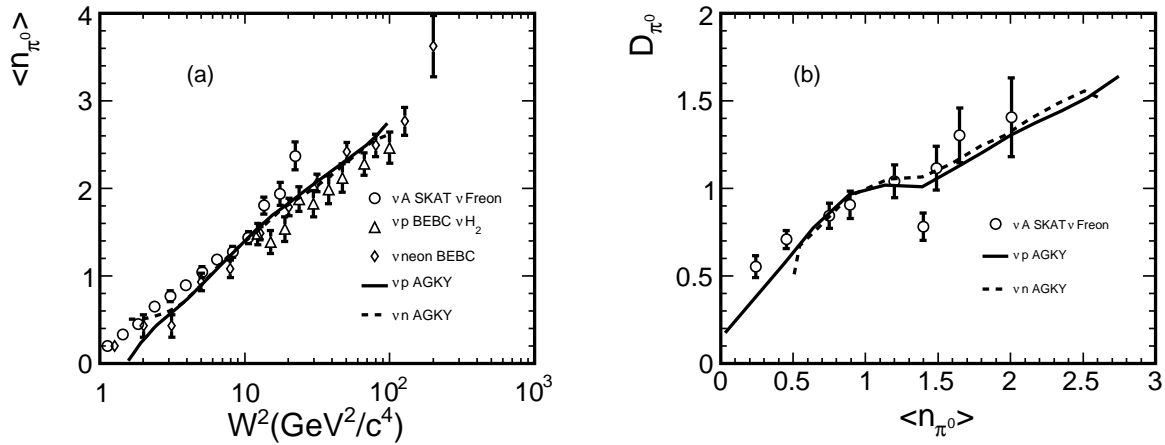}
  \caption{(a) Average multiplicity of $\pi^{0}$ mesons as a function of $W^{2}$. (b) Dispersion of the distributions in multiplicity as a function of the average multiplicity of $\pi^{0}$ mesons. Data points are taken from \cite{Wittek:1988ke,Ivanilov:1984gh,Grassler:1983ks}}\label{fig:cMulPi0}
\end{figure*}

\begin{figure*}
\centering
  \includegraphics[width=\textwidth]{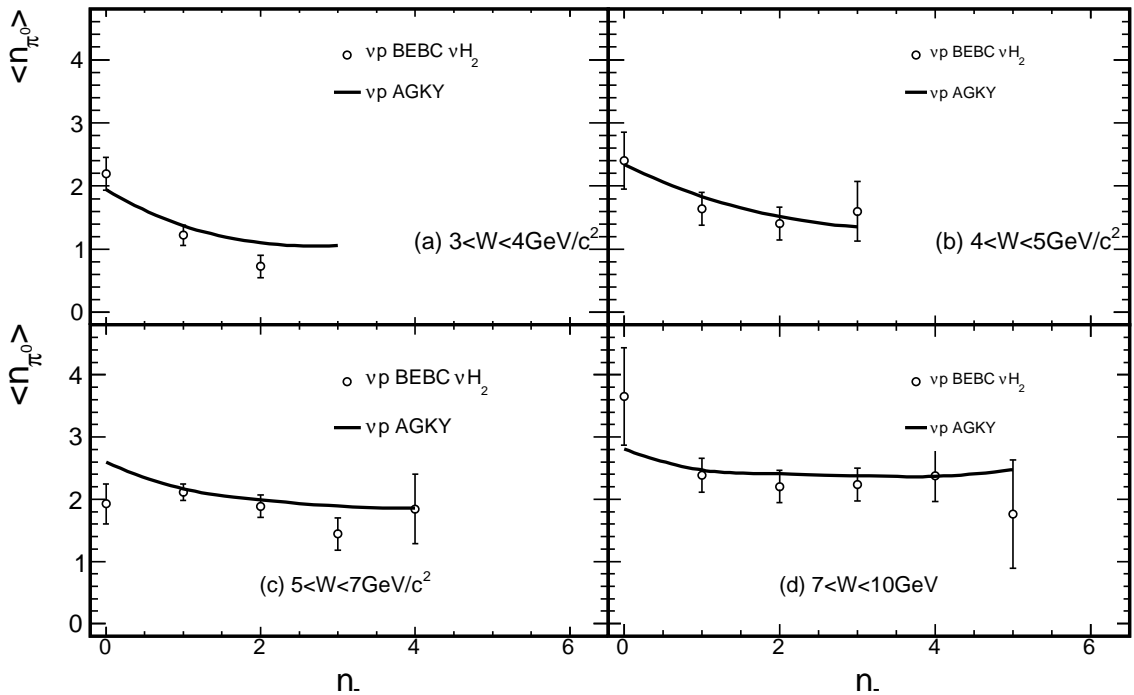}
  \caption{Average $\pi^{0}$ multiplicity $\langle n_{\pi^{0}} \rangle$ as a function of the number of negative hadrons $n_{-}$ for different intervals of $W$. Data points are taken from \cite{Grassler:1983ks}.}\label{fig:cCorr_Pi0_Ch}
\end{figure*}

Fig.\ref{fig:cMulCh} shows the average charged hadron multiplicity $\langle n_{ch} \rangle$ (the number of charged hadrons in the final state, {\it i.e.} excluding the muon) as a function of $W^{2}$. $\langle n_{ch} \rangle$ rises linearly with $\ln(W^{2})$ for $W>2 \mbox{GeV}/c^{2}$. At the lowest $W$ values the dominant interaction channels 
are single pion production from baryon resonances:
\begin{eqnarray}
\label{res1}
\nu+p&\rightarrow&\mu^{-}+p+\pi^{+}\\
\label{res2}
\nu+n&\rightarrow&\mu^{-}+p+\pi^{0}\\
\label{res3}
\nu+n&\rightarrow&\mu^{-}+n+\pi^{+}
\end{eqnarray}
Therefore $\langle n_{ch} \rangle$ becomes 2(1) for $\nu p$($\nu n$) interactions as $W$ approaches the pion production threshold. For $\nu p$ interactions there is a disagreement between the two measurements especially at high invariant masses, which is probably due 
to differences in scattering from hydrogen and deuterium targets.     
Our parameterization of low-$W$ model was based on the Fermilab 15-foot chamber data. Historically the PYTHIA/JETSET program was tuned on the BEBC data. The AGKY model uses the KNO-based empirical model at low invariant masses and it uses the PYTHIA/JETSET program to simulation high invariance mass interactions. Therefore the MC prediction agrees better with the Fermilab data at low invariant masses and it agrees better with the BEBC data at high invariant masses.

\begin{figure*}
\centering
  \includegraphics[width=\textwidth]{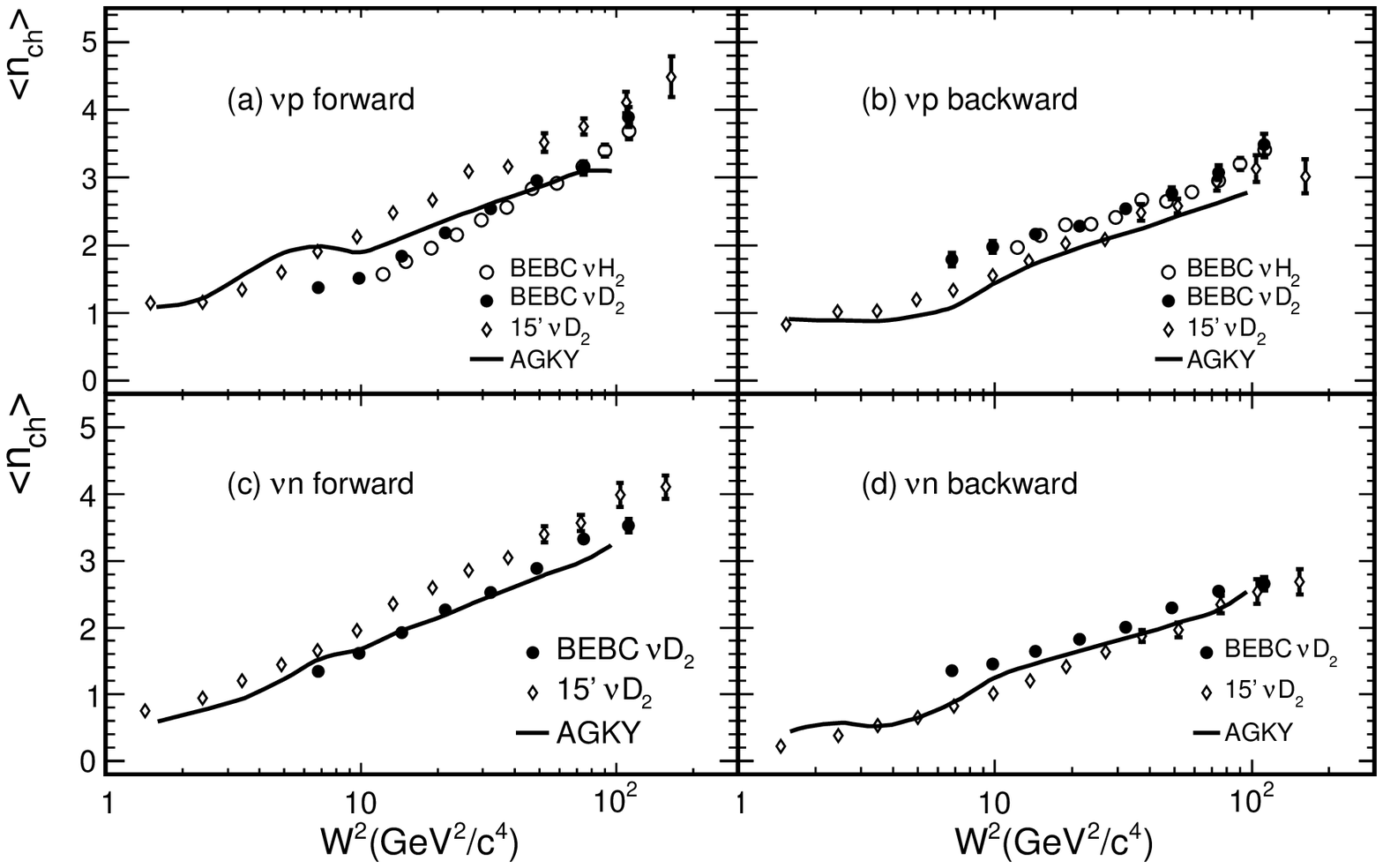}
  \caption{Average charged-hadron multiplicity in the forward and backward hemispheres as functions of $W^{2}$: (a) $\nu p$, forward, (b) $\nu p$, backward, (c) $\nu n$, forward, (d) $\nu n$, backward. Data points are taken from \cite{Zieminska:1983bs,Grassler:1983ks,Allasia:1984ua}.}\label{fig:cMulFB}
\end{figure*}

The production of strange particles via associated production is shown in Figures \ref{fig:kaons}
and \ref{fig:lambdas}.  The production of kaons and lambdas for the KNO-based model are in 
reasonable agreement with the data, while the rate of strange meson production from JETSET is 
clearly low.  We have investigated adjusting JETSET parameters to produce better agreement with 
data.  While it is possible to improve the agreement with strange particle production data, 
doing so yields reduced agreement with other important distributions, such as the normalized 
charged particle distributions.  

\begin{figure}
\centering
    \includegraphics[width=0.5\textwidth]{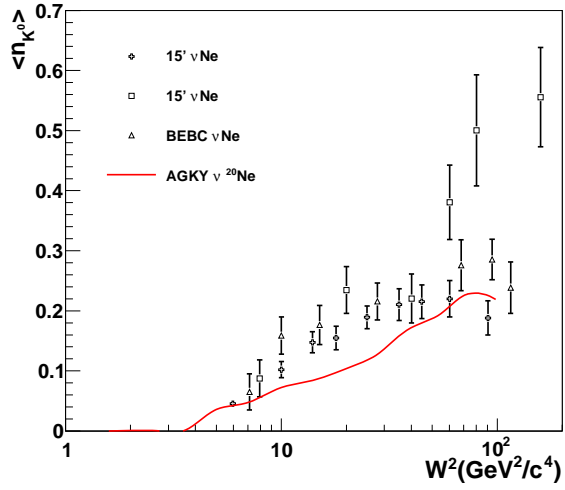}
\caption{Neutral kaon production rate on neon as a function of invariant mass.   
Data are from \cite{Bosetti:1982vk,Baker:1986xx,DeProspo:1994ac}. \label{fig:kaons}}
\end{figure}

\begin{figure}
\centering
    \includegraphics[width=0.5\textwidth]{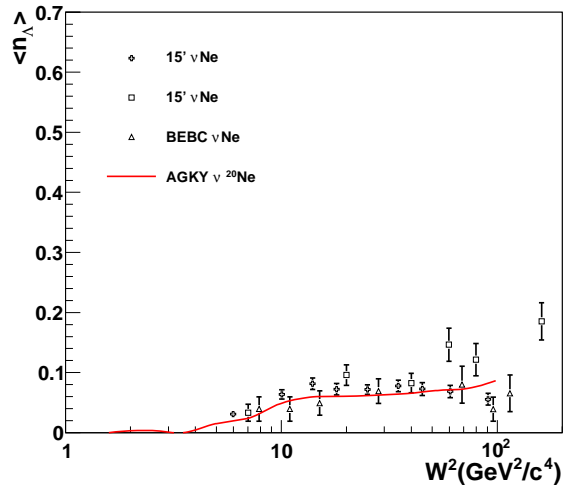}
\caption{Lambda production rate as a function of invariant mass. 
Data are from \cite{Bosetti:1982vk,Baker:1986xx,DeProspo:1994ac}. \label{fig:lambdas}}
\end{figure}

Fig.\ref{fig:cDispCh}(a) shows the dispersion $D_{-} = (\langle n_{-}^{2} \rangle - \langle n_{-} \rangle ^{2})^{1/2}$ of the negative hadron multiplicity as a function of $\langle n_{-} \rangle$. Fig.\ref{fig:cDispCh}(b) shows the ratio $D/\langle n_{ch} \rangle$ as a function of $W^{2}$. The dispersion is solely determined by the KNO scaling distributions shown in Fig.\ref{fig:kno_levy}. The agreement between data and MC predictions is satisfactory. 

\begin{figure*}
\centering
  \includegraphics[width=\textwidth]{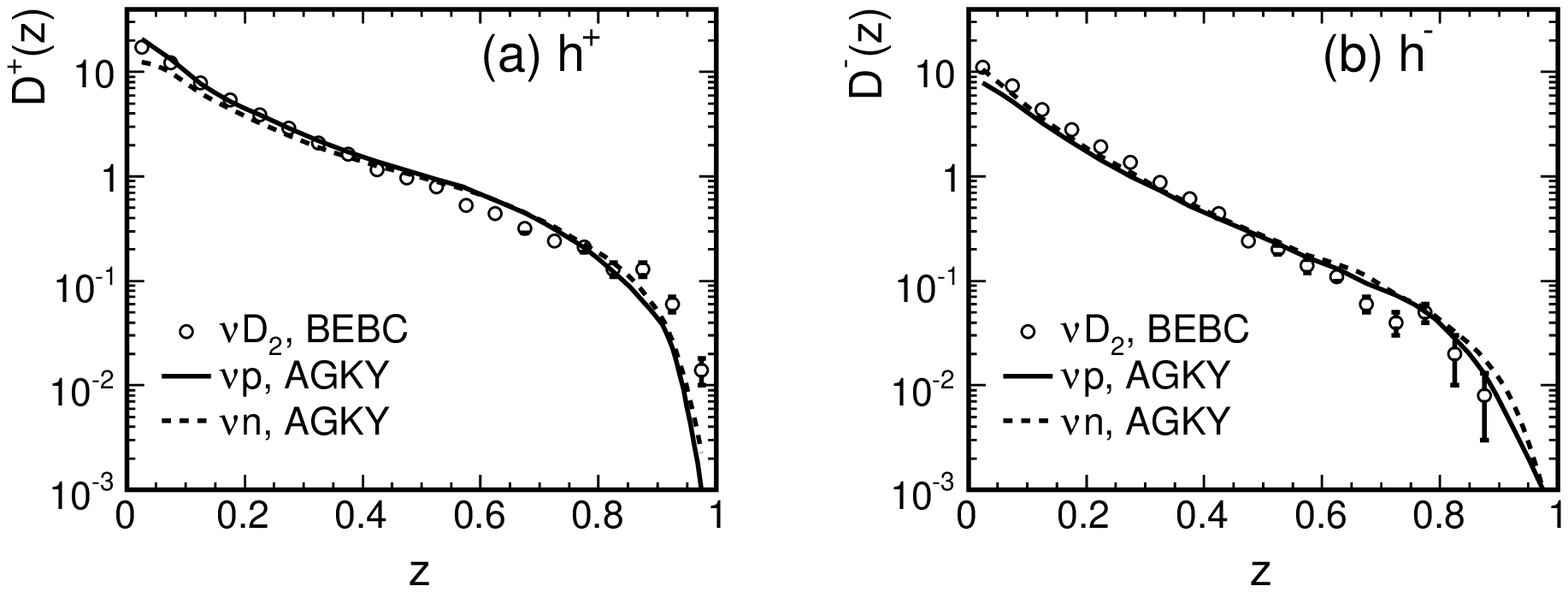}
  \caption{Fragmentation functions for positive (a) and negative (b) hadrons. Applied cuts: $W^{2} > 5 (GeV/c^{2})^{2}$, $Q^{2} > 1 (GeV/c)^{2}$. Data points are taken from \cite{Allasia:1984ua}.}\label{fig:cZ}
\end{figure*}

Fig.\ref{fig:cMulPi0}(a) shows the average $\pi^{0}$ multiplicity $\langle n_{\pi^{0}} \rangle$ as a function of $W^{2}$. Fig.\ref{fig:cMulPi0}(b) shows the dispersion of the distributions in multiplicity as a function of the average multiplicity of $\pi^{0}$ mesons. As we mentioned it is difficult to detect $\pi^{0}$'s inside a hydrogen bubble chamber. Also shown in the plot are some measurements using heavy liquids such as neon and Freon. 
In principle, rescattering of the primary hadrons can occur in the nucleus. Some studies of inclusive negative hadron production in the hydrogen-neon mixture and 
comparison with data obtained by using hydrogen targets indicate that these effects are negligible \cite{Berge:1978fr}. 
The model is in good agreement with the data. $\langle n_{\pi^{0}} \rangle$ is 0(1/2) for $\nu p$($\nu n$) interactions when the hadronic invariant mass approaches the pion production threshold, which is consistent with the expectation from the reactions (\ref{res1}-\ref{res3}). The model predicts the same average $\pi^{0}$ multiplicity for $\nu p$ and $\nu n$ interactions for $W>2\mbox{GeV}/c^{2}$.

Fig.\ref{fig:cCorr_Pi0_Ch} shows the average $\pi^{0}$ multiplicities $\langle n_{\pi^{0}} \rangle$ as a function of the number of negative hadrons $n_{-}$ for various $W$ ranges. At lower $W$, $\langle n_{\pi^{0}} \rangle$ tends to decrease with $n_{-}$, probably because of limited phase space, while at higher $W$ $\langle n_{\pi^{0}} \rangle$ is rather independent of $n_{-}$ where there is enough phase space.    
Our model reproduces the correlation at lower $W$ suggested by the data. However, another experiment measured the same correlation using neon-hydrogen mixture and their results indicate that $\langle n_{\pi^{0}} \rangle$ is rather independent of $n_{-}$ for both $W>4\mbox{GeV}/c^{2}$ and $W<4\mbox{GeV}/c^{2}$ \cite{Ammosov:1978vt}. Since events with $\pi^{0}$ but with 0 or very few charged pions are dominant background events in the $\nu_{e}$ appearance analysis, it is very important to understand the correlation between the neutral pions and charged pions; this should be a goal of future experiments \cite{Drakoulakos:2004gn}.   

\begin{figure*}
\centering
  \includegraphics[width=\textwidth]{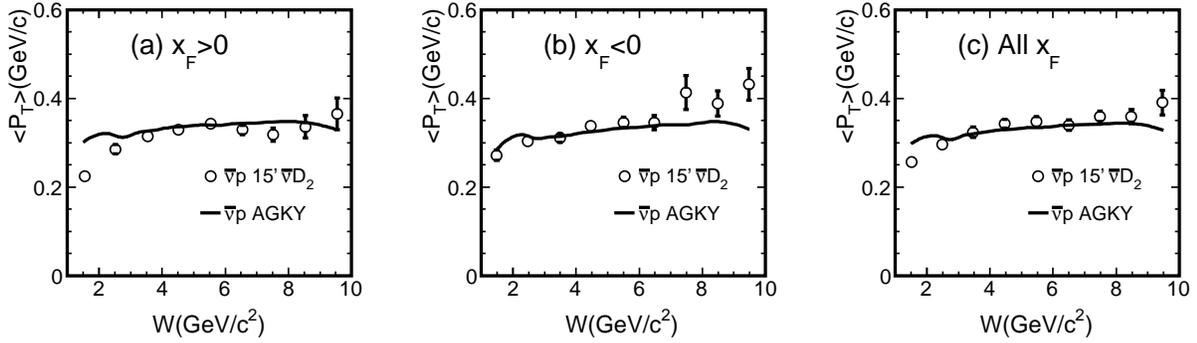}
  \caption{Mean value of the transverse momentum of charged hadrons as a function of $W$ for the selections (a) $x_{F}>0$, (b) $x_{F}<0$, and (c) all $x_{F}$. Data points are taken from \cite{Derrick:1981br}.}\label{fig:cPt_W}
\end{figure*}

Fig.\ref{fig:cMulFB} shows the average charged-hadron multiplicity in the forward and backward hemispheres as functions of $W^{2}$. 
The forward hemisphere is defined by the direction of the current in the total hadronic c.m.s. 
There is a bump in the MC prediction in the forward hemisphere for $\nu p$ interactions at $W\sim2\mbox{GeV}/c^{2}$ and there is a slight dip in the backward hemisphere in the same region. This indicates that the MC may overestimate the hadrons going forward in the hadronic c.m.s. at $W\sim2\mbox{GeV}/c^{2}$ and underestimate the hadrons going backward. One consequence could be that the MC overestimates the energetic hadrons since the hadrons in the forward hemisphere of hadronic c.m.s. get more Lorentz boost than those in the backward hemisphere when boosted to the LAB frame. This may be caused by the way we determine the baryon 4-momentum and preferably select events with low $p_{T}$ in the phase space decay.   These effects will be investigated further for improvement in future versions of the model. 

Fig.\ref{fig:cZ} shows the fragmentation functions for positive and negative hadrons. The fragmentation function is defined as: $D(z) = \frac{1}{N_{ev}}\cdot\frac{dN}{dz}$, where $N_{ev}$ is the total number of interactions (events) and $z=E/\nu$ is the fraction of the total energy transfer carried by each final hadron in the laboratory frame. The AGKY predictions are in excellent agreement with the data.

Fig.\ref{fig:cPt_W} shows the mean value of the transverse momentum with respect to the current direction of charged hadrons as a function of $W$. The MC predictions match the data reasonably well. In the naive QPM, the quarks have no transverse momentum within the struck nucleon, and the fragments acquire a $P_{T}^{frag}$ with respect to the struck quark from the hadronization process. The average transverse momentum $\langle P_{T}^{2} \rangle$ of the hadrons will then be independent of variables such as $x_{BJ}$, $y$, $Q^{2}$, $W$, {\it etc.}, apart from trivial kinematic constraints and any instrumental effects. Both MC and data reflect this feature. However, in a perturbative QCD picture, the quark acquires an additional transverse component, $\langle P_{T}^{2} \rangle^{QCD}$, as a result of gluon radiation.  
The quark itself may also have a primordial $\langle P_{T}^{2} \rangle^{prim}$ inside the nucleon. 
These QCD effects can introduce dependencies of $\langle P_{T}^{2} \rangle$ on the variables $x_{BJ}$, $y$, $Q^{2}$, $W$, $z$, {\it etc.}

Fig.\ref{fig:cPt_xf} shows the mean value of the transverse momentum of charged hadrons as a function of $x_{F}$, where $x_{F} = \frac{p_{L}^{*}}{p_{Lmax}^{*}}$ is the Feynman-x.  As is well known, $\langle p_{T} \rangle$ increases with increasing $|x_{F}|$ with a shape called the seagull effect. This effect is reasonably well modeled by the AGKY model. 

\begin{figure*}
\centering
  \includegraphics[width=\textwidth]{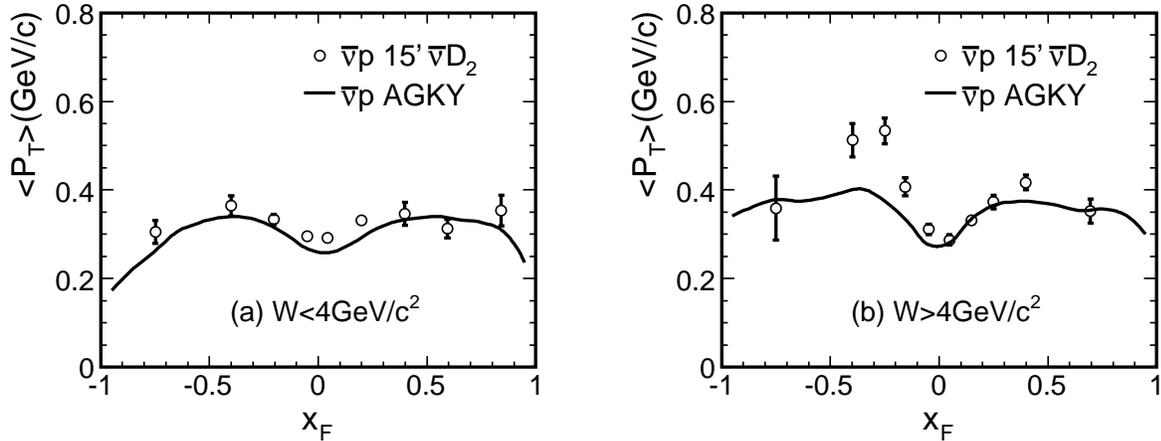}
  \caption{Mean value of the transverse momentum of charged hadrons as a function of $x_{F}$ for $\bar{\nu} p$. (a) $W<4\mbox{GeV}/c^{2}$, (b) $W>4\mbox{GeV}/c^{2}$. Data points are taken from \cite{Derrick:1981br}.}\label{fig:cPt_xf}
\end{figure*}

\section{Conclusions}
\label{sec:7}

In this paper we have described a new hadronic mutiparticle production model for use in neutrino simulations.   This model 
will be useful for experiments in the few-GeV energy regime and exhibits satisfactory agreement with wide variety of data for 
charged, neutral pions as well as strange particles.  Several upcoming expriments will have high-statistics data sets in detectors
with excellent energy resolution, neutral particle containment, and particle identification.  These experiments are in some cases
considering possible running with cryogenic hydrogen and deuterium targets.     These experiments will be operating in 
this few-GeV regime and have the potential to fill in several gaps in our understanding that will help improve hadronic shower modeling
for oscillation experiments.  

The upcoming generation of experiments have all the necessary prerequisites to significantly address the existing experimental uncertainties
in hadronization at low invariant mass.   These result from the fact that these detectors have good containment for both charged and neutral 
particles, high event rates, good tracking resolution, excellent particle identification and energy resolution, and the 
possibility of collecting data on free nucleons with cryogenic targets.  
The latter offers the possibility of addressing the challenge of disentangling hadronization modeling from intranuclear rescattering effects. 
Charged current measurements of particular interest 
will include clarifying the experimental discrepancy at low invariant mass between the existing published results as shown in 
Fig.\ref{fig:cMulFB}, the origin of which probably relates to particle misidentification corrections \cite{Grassler:1983ks}.  This discrepancy 
has a large effect on forward/backward measurements, and a succesful resolution of this question 
will reduce systematic differences between datasets in a large class
of existing measurements.  In addition, measurements of transverse momentum at low invariant masses will be helpful in model tuning. 
Measurements of neutral particles, in particular multiplicity and particle dispersion from free targets at low invariant mass, will 
be tremendously helpful.  The correlation between neutral and charged particle multiplicities at low invariant  mass
is particularly important for 
oscillation simulations, as it determines the likelihood that a low invariant mass shower will be dominated by neutral pions. 

\section{Acknowledgements}
The authors would like to thank W.A. Mann, J. Morfin, and S. Wojcicki for helpful comments and discussions.  This work was supported by 
Department of Energy grant DE-FG02-92ER40702 and the Tufts Summer Scholars program. 

\bibliographystyle{epj}
\bibliography{AGKY3}

\end{document}